\documentclass[12pt]{iopart}

\usepackage{amssymb}
\usepackage{graphicx,color}
\usepackage{bm}
\usepackage{color}
\usepackage{amsthm}

\usepackage{iopams}

\newcommand{\ToDelete}[1]{}
\newcommand{\ToAdd}[1]{#1}

\newcommand{\Ket}[1]{\left\vert #1\right\rangle}

\newcommand{\KetBra}[2]{\left\vert#1\right\rangle\left\langle#2\right\vert}

\newcommand{\Projector}[1]{\KetBra{#1}{#1}}
\def\ii{\mathrm{i}}

\newcommand{\eqref}[1]{Eq.~(\ref{#1})}
\newcommand{\text}[1]{\mathrm{#1}}

\begin{document}

\title{Zeno Dynamics and High-Temperature Master Equations Beyond Secular
Approximation}

\author{B. Militello $^1$, M. Scala $^2$, A. Messina $^1$}

\address{ $^1$ Dipartimento di Fisica e Chimica, Universit\`{a} degli
Studi di Palermo, Via Archirafi 36, 90123, Palermo. Italy}
\address{ $^2$ Department of Physics and Astronomy, University
College London, Gower Street, London WC1E 6BT, United Kingdom}

\begin{abstract}
Complete positivity of a class of maps generated by master
equations derived beyond the secular approximation is discussed.
The connection between such class of evolutions and physical
properties of the system is analyzed in depth. It is also shown
that under suitable hypotheses a Zeno dynamics can be induced
because of the high temperature of the bath.
\end{abstract}



\maketitle

\section{Introduction}

Quantum Zeno effect (QZE) is an ubiquitous phenomenon in quantum
mechanics \cite{ref:MishraSudarshan, ref:PascazioFacchi2004,
ref:PascazioFacchi2008, ref:PascazioFacchi2010}, relevant in the
analysis of fundamental concepts of quantum mechanics
\cite{ref:Home1997,ref:Schulman1998,ref:Panov1999a,ref:Panov1999b},
useful in applications \cite{ref:Applications}, and which has been
experimentally demonstrated
\cite{ref:Itano1990,ref:Fischer1997,ref:Fischer2001}.

Recently, the connection between occurrence of Zeno phenomena and
the properties of the environment of the physical system under
scrutiny has been investigated \cite{ref:Kofman2001,
ref:Ruseckas2002, ref:Kavan2007, ref:Sabrina2006}. The connection
between QZE and high-temperature has been studied. Inhibition of
Landau-Zener transitions induced by temperature has been predicted
\cite{ref:Scala_LZ}, and enhancement of QZE phenomena through
temperature has been analyzed in depth
\cite{ref:Militello_ThermalQZE}. In the case of high temperature,
exploitation of \ToDelete{Time Convolution-less (TCL)} \ToAdd{the
standard} approach to derive master equations, and consequent
secular approximation could be problematic, due to the fact that
in such a case the secular approximation, generally performed
after Born-Markov approximation, can be inappropriate because of
an effectively high coupling strength induced by the increased
number of photons. Nevertheless, in general, avoiding secular
approximation would produce a generator of a non-completely
positive map. Since complete positivity is a necessary condition
for a map describing a physical system
\cite{ref:Complete_Positivity}, one needs to overcome such
difficulty. Studies of quantum optics models obtained beyond the
secular approximation or without the rotating wave approximation
(two names used in different contexts to address essentially the
same approximation) have been reported through the years in
different papers \cite{ref:BeyondSecularApprox}.

In this paper, we find suitable hypotheses under which
\ToDelete{TCL} \ToAdd{Markovian} approach, in the limit of high
temperature, can produce a completely positive map even beyond the
secular approximation. Moreover, through the master equation
derived along this route, it is possible to forecast the
occurrence of a Zeno phenomenon under suitable hypotheses. This
outcome somehow confirms the results obtained in
\cite{ref:Scala_LZ} and \cite{ref:Militello_ThermalQZE}. The paper
is organized as follows. In the next section we introduce the
general approach and assumptions to derive the master equation in
the high temperature limit beyond the secular approximation. In
section \ref{sec:3_level_system} we provide a general statement
connected with the appearance of Zeno phenomena, and show an
example in a three-level system where the high temperature master
equation clearly forecasts preservation of a subspace, which we
address as thermal quantum Zeno effect. Finally, in the last
section, we give some conclusive remarks.

\section{Master Equation}\label{sec:general_scheme}

Let us consider a system interacting with a bosonic environment:
\begin{eqnarray}
H_{\text{S}} &=& \sum_\epsilon \epsilon \Pi (\epsilon)\,,\\
H_{\text{B}} &=& \sum_k \omega_k a^\dag_k a_k\,,\\
H_{\text{I}} &=& A \otimes B\,, \qquad %
B =\sum_{k}\left(g_k^* a_k+g_k a_k^\dag\right)\,,
\end{eqnarray}
where $\Pi (\epsilon)$ are projectors on eigenspaces of
$H_{\text{S}}$ \ToAdd{and $A=A^\dag$}.

\ToDelete{Now, following the standard approach
\cite{ref:Petruccione, ref:Gardiner}, we write down the iterated
von Neumann equation, after performing the Born-Markov
approximation}

\ToAdd{

Now, following the standard approach \cite{ref:Petruccione,
ref:Gardiner}, in the weak coupling limit, we can assume that the
total density matrix can be separated in the system and bath
degrees of freedom (Born approximation): $\rho_\mathrm{T} = \rho
\otimes \rho_\mathrm{B}$, where $\rho$ is the density operator of
the system $\mathrm{S}$ and $\rho_\mathrm{B}$ is the thermal state
of the bosonic bath --- the state of the bath does not
significantly change, both because of the weak coupling limit and
the fact that the environment is much larger than the system and
therefore negligibly affected by it. Then we formally integrate
the von Neumann equation, iterate it and differentiate both
members of the relation obtained:
\begin{equation}\label{eq:IteratedVonNeumann}
\frac{\mathrm{d}\rho(t)}{\mathrm{d}t} = - \int_0^t \, \mathrm{d}s
\, \mathrm{tr}_\mathrm{B} \, [H_I(t), [H_I(s),
\rho(s)\otimes\rho_\mathrm{B}]]\,,
\end{equation}
where, considered the structure of the operator $B$ and of the
thermal state $\rho_\mathrm{B}$, one has
$\mathrm{tr}_\mathrm{B}[\rho_\mathrm{B}, B] = 0$.

Subsequently, we perform the Markov approximation by substituting
the density operator $\rho(s)$ in the integral with $\rho(t)$, and
then (by assuming very short correlation time for the bath) we put
the second limit of integration to infinity: }
\begin{equation}\label{eq:BornMarkovApprox}
\frac{\mathrm{d}\rho(t)}{\mathrm{d}t} = - \int_0^\infty \,
\mathrm{d}s \, \mathrm{tr}_\mathrm{B} \, [H_I(t),
[H_I(\ToAdd{t-}s),
\rho(t)\otimes\rho_\mathrm{B}]]\,. %
\end{equation}
Then we introduce the jump operators:
\begin{eqnarray}
A(\omega) &= \sum_{\epsilon' - \epsilon = \omega} \Pi(\epsilon) \,
A \, \Pi(\epsilon')\,,
\end{eqnarray}
where the sum is meant over all $\epsilon$ and $\epsilon'$ such
that $\epsilon' - \epsilon = \omega$. By exploiting such operators
and before performing the secular approximation, we obtain:
\begin{eqnarray}\label{eq:MasterEq}
\nonumber \frac{\mathrm{d}\rho}{\mathrm{d}t} &=&
-i\left[H_{\text{S}}, \rho \right] + \sum_{\omega,\,\omega'}\,\,\,
\Gamma(\omega) \left[A(\omega) \rho\, A^\dag(\omega') \ToAdd{-}
A^\dag(\omega') A(\omega) \rho \right] +
\mathrm{H.c.}\,,
\end{eqnarray}
with $\omega$ and $\omega'$ spanning over all possible Bohr
frequencies of the system, and
\begin{eqnarray}
\nonumber \Gamma(\omega) &=& \int_0^\infty \mathrm{d}s\,e^{i\omega
s} \mathrm{tr}_B\left[B^\dag(t) B(t-s) \rho_B(0)\right]\,\\
&=& \nonumber \left\{%
\begin{array}{ll}
\left|g(\omega)\right|^2 D(\omega)
(1+N(\omega))\,, \qquad & \omega > 0\\
& \\
\left|g(|\omega|)\right|^2 D(|\omega|) \, N(|\omega|)\,, \qquad &
\omega < 0\,,
\end{array}
\right.\\
\end{eqnarray}
where $D(\omega)$ is the bath density of modes, while $g(\omega)$
is the coupling constant $g_k$ in the continuum limit, and
$N(\omega) = 1/(\exp(\hbar\omega/(k_B T)-1)$. In the derivation of
Eq.(\ref{eq:MasterEq}) we have neglected the Lamb shifts (LS),
which, on the other end, if not negligible can be absorbed by the
commutator after introducing $\tilde{H}_{\text{S}} = H_{\text{S}}
+ \text{LS}$ and $[H_{\text{S}}, \rho ] \rightarrow
[\tilde{H}_{\text{S}}, \rho ]$.

Generally, Eq.(\ref{eq:MasterEq}) does not define the generator of
a completely positive map. What one usually does is to perform the
secular approximation in order to find a completely positive
generator. This implies keeping only the diagonal terms
($\omega=\omega'$) in \eqref{eq:MasterEq} and neglecting the rest.

Anyway, as we are going to demonstrate, there exist suitable
hypotheses under which the structure of the \ToDelete{TCL}
\ToAdd{Markovian} master equation without secular approximation
approaches a generator of a CP map.

\ToDelete{Assume that the operator $A$ connects only two
subspaces, $\Pi_1$ and $\Pi_2$ generated by projectors on
eigenspaces of $H_{\text{S}}$, which means:}

\begin{figure}
\centering
\includegraphics[width=0.35\textwidth, angle=0]{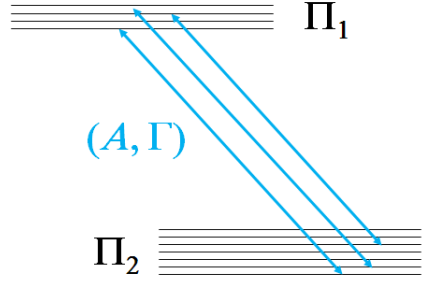} %
\caption{(Color online). Assumptions on the coupling scheme: the
two subspaces $\Pi_1$ and $\Pi_2$ are connected through the
operator $A$,
but no transition is induced by $A$ within each subspace.}%
\label{fig:general_scheme}
\end{figure}

\ToAdd{Assume that there are two subspaces, say $1$ and $2$,
corresponding to two projectors $\Pi_1$ and $\Pi_2$ generated by
projectors on eigenspaces of $H_{\text{S}}$. Assume also that the
operator $A$ connects only such two subspaces, which means:}
\begin{eqnarray}
  \label{eq:A_Diagonal}
  &&
  \Pi_1 \, A \, \Pi_1 = 0 \,, \qquad
  \Pi_2 \, A \, \Pi_2 = 0 \,,\\
  \label{eq:A_OffDiagonal}
  &&
  \Pi_1 \, A \, \Pi_2 \not= 0 \,, \qquad
  \Pi_2 \, A \, \Pi_1 \not= 0\,.
\end{eqnarray}
Eqs.(\ref{eq:A_Diagonal}) express the fact that $A$ does not
induce transitions within each of the two subspaces (all matrix
elements involving any couple of states of the same subspace are
vanishing). Transitions from one subspace to the other are instead
allowed, according to eqs.(\ref{eq:A_OffDiagonal}). In
Fig.~\ref{fig:general_scheme} we provide a graphical
representation of this scenario. If $\omega$ refers to a
transition within $\Pi_1$ or within $\Pi_2$, then the
corresponding $A(\omega)$ vanishes, due to
Eq.(\ref{eq:A_Diagonal}). Instead, if $\omega$ refers to
transitions between the two subspaces then the corresponding
$A(\omega)$ is non-vanishing. All this implies that in
Eq.(\ref{eq:MasterEq}) summation can be restricted to $\omega$ and
$\omega'$ lying in the inter-band transition frequencies
(frequencies related to transitions from $\Pi_1$ to $\Pi_2$ and
vice versa). Let us introduce $\omega_0$ as the frequency
connecting the centre of the band associated to $\Pi_1$ with the
centre of the band associated to $\Pi_2$. We will use the notation
$\sum_{\omega\sim\omega_0}$ to indicate summation over values of
$\omega$ (whose absolute values are) close to $\omega_0$.

Assume also that the Bohr frequencies related to transitions from
any state in $\Pi_1$ to any state in $\Pi_2$ are much larger than
the Bohr frequencies associated either to transitions inside
$\Pi_1$ or to transitions inside $\Pi_2$.

Finally, assume that the spectrum is flat, meaning that
$\left|g(|\omega|)\right|^2 D(|\omega|) = \gamma$ is independent
of $\omega$, at least in a neighborhood of $\omega_0$
corresponding to the frequencies of interest for this physical
problem. Of course, $\Gamma(\omega)$ still contains a dependence
on $\omega$ due to the term $N(\omega)$.

Now, in the limit of high-temperature (which implies $N(\omega)
\gg 1$) and weak coupling ($\gamma \rightarrow 0$), we can
consider $\gamma(1+N(\omega)) \approx \gamma N(\omega)$. Moreover,
\ToAdd{in the high-temperature limit} it turns out $N(\omega)
\approx k_B T / (\hbar\omega)$, and since the frequencies
connecting the subspaces $1$ and $2$ are very close
--- they are all much larger than the Bohr frequencies associated
to transitions internal to each subspace --- then one can deduce
that $N(\omega)$ does not significantly depend on $\omega$, when
$\omega$ is related to a transition between the two subspaces. In
particular, one has ($\omega = \omega_0 + \delta\omega$)
\begin{eqnarray}
\nonumber N(\omega) &\approx & N(\omega_0) + \left.\frac{\partial
N}{\partial\omega}\right|_{\omega=\omega_0} \delta\omega\\
\nonumber &=& N(\omega_0) -
[N(\omega_0)]^2\,\exp\left({\frac{\hbar\omega_0}{k_\mathrm{B}T}}\right)
\frac{\hbar\delta\omega}{k_\mathrm{B}T} \\
&\approx & N(\omega_0) -
\frac{k_\mathrm{B}T\delta\omega}{\hbar\omega_0^2} \,,
\end{eqnarray}
where the last step is legitimated at high temperature, since in
this limit $N(\omega_0) \approx k_\mathrm{B}T/(\hbar\omega_0)$ and
the exponential $\exp(\hbar\omega_0/(k_\mathrm{B}T))$ approaches
unity.

Therefore, Eq.(\ref{eq:MasterEq}) may be cast in the following
form:
\begin{eqnarray}\label{eq:MasterEq_simplified}
\nonumber \frac{\mathrm{d}\rho}{\mathrm{d}t} &\approx
-i\left[H_{\text{S}}, \rho \right] + \gamma \left\{ %
\sum_{\omega,\,\omega' \sim \omega_0}\,\,\,  N(\omega)
\left[A(\omega)\rho\, A^\dag(\omega') - A^\dag(\omega') A(\omega)
\rho \right] + \mathrm{H.c.}
\right\} \,
\\ %
\nonumber
&\approx %
-i\left[H_{\text{S}}, \rho \right] + \gamma N(\omega_0) %
\left\{ %
\sum_{\omega,\,\omega' \sim \omega_0}\,\,\, [A(\omega)\rho\,
A^\dag(\omega') - A^\dag(\omega') A(\omega) \rho ] + \mathrm{H.c.}
\right\} \\
&+
\ToAdd{\gamma\, O(k_\mathrm{B}T\delta\omega/(\hbar\omega_0^2))} %
\,.
\end{eqnarray}

Since $A = \sum_\omega A(\omega) =$ $\sum_{\omega \sim \omega_0}
A(\omega) =$ $\sum_{\omega \sim \omega_0} A^\dag(\omega) =$
$\sum_\omega A^\dag(\omega) =$ $A^\dag$,
\eqref{eq:MasterEq_simplified} can be put in the form:
\begin{eqnarray}
\nonumber \frac{\mathrm{d}\rho}{\mathrm{d}t} &=&
-i\left[H_{\text{S}}, \rho \right] + \gamma N(\omega_0) %
\left[ %
( A\,\rho\,A - A^2\,\rho ) + \mathrm{H.c.} \right]\\
&+& \gamma O( k_B T \, \delta\omega/(\hbar\omega_0^2)) %
\,,
\end{eqnarray}
which, in the limit $\gamma k_B T \delta\omega / (\hbar\omega_0^2)
\rightarrow 0$, becomes,
\begin{equation}\label{eq:MasterEq_simplified_more}
\frac{\mathrm{d}\rho}{\mathrm{d}t} = -i\left[H_{\text{S}}, \rho
\right] + \gamma N(\omega_0) %
\left[ %
( 2 A\,\rho\,A - A^2\,\rho - \rho\,A^2 ) \right] %
\,.
\end{equation}

Since this master equation is in the standard form, the relevant
evolution is described by a completely positive map.

Two points are noteworthy: the complete positivity of the induced
map (even considered the derivation beyond the secular
approximation), and the independence of the dissipator from the
specific form of $H_{\text{S}}$. This second property is related
to the fact that at a certain point we sum up over all the jump
operators ($A(\omega)$), which \lq reconstructs\rq\, the complete
system operator involved in the system-bath interaction ($A$).

\ToDelete{It is also interesting to sum up the hypotheses we have
exploited: weak coupling ($\gamma \rightarrow 0$), high
temperature (which implies $N(\omega) \gg 1$) and band structure
for the energy spectrum of the system ($\delta\omega/\omega_0
\rightarrow 0$). These three different limits should be taken in
such a way that $\gamma k_B T \delta\omega / (\hbar\omega_0^2)
\rightarrow 0$, which can be realized in many different ways.}

\ToAdd{It is also interesting to sum up the hypotheses we have
exploited. The first one is the typical weak coupling limit
($\gamma \rightarrow 0$) which is important to perform the Born
approximation and start with the derivation of the master
equation. The second hypothesis is the high temperature limit
(which implies $N(\omega) \gg 1$) which allows to consider
essentially equal the rates of the downward and upward transitions
induced by the environment. The weak coupling limit makes even
closer such rates, for any given frequency. Finally, the band
structure for the energy spectrum of the system (also assuming
that inter-band transitions are forbidden) and the narrowness of
such bands ($\delta\omega/\omega_0 \rightarrow 0$) allow to
consider equal the rates for different frequencies, making it
possible to reconstruct the operator $A$ from the relevant jump
operators. These three different limits should be taken in such a
way that $\gamma k_B T \delta\omega / (\hbar\omega_0^2)
\rightarrow 0$ (see Eq.~(\ref{eq:MasterEq_simplified})), which can
be realized in many different ways. In particular, a weaker
system-environment coupling or narrower bands allow to explore the
thermal Zeno phenomenon at higher temperature.}

\section{Thermal Zeno Dynamics}\label{sec:3_level_system}

{\bf\it General Statement --- } The structure of the master
equation in \eqref{eq:MasterEq_simplified_more} suggests the idea
that under suitable hypotheses the dissipator can hinder some
dynamical effects of $H_{\text{S}}$, the free Hamiltonian of the
system.

This statement can be proven through \ToDelete{the following}
considerations \ToAdd{similar to those made in previous works, in
Hamiltonian
contexts~\cite{ref:PascazioFacchi2004,ref:PascazioFacchi2010} and
in dissipative contexts~\cite{ref:Scala_LZ}}. Since $\alpha \equiv
\gamma N(\omega_0)$ is very large, we can treat the commutator
$-i\left[H_{\text{S}}, \rho \right]$ as a perturbation with
respect to the dissipator. Assuming this point of view, it is easy
to convince oneself that the Hamiltonian cannot significantly
connect subspaces of the dissipator which are well separated in
terms of the relevant eigenvalues. In other words, assuming that
$\rho_{\text{a(b)}}$ is an eigenoperator of the dissipator
associated to the eigenvalue $\alpha\lambda_{\text{a(b)}}$ ---
this means that $2 A \rho_{\text{a(b)}} A - A^2\rho_{\text{a(b)}}
- \rho_{\text{a(b)}}A^2 = \lambda_{\text{a(b)}}
\rho_{\text{a(b)}}$
--- then, provided $\alpha
|\lambda_{\text{a}}-\lambda_{\text{b}}|$ is much larger than the
matrix elements of $H_{\text{S}}$, the action of the commutator
can only weakly connect the operators $\rho_{\text{a}}$ and
$\rho_{\text{b}}$. Now, by increasing the temperature (and then
$\alpha$) the eigenvalues and their differences increase, which
makes more and more ineffective the free Hamiltonian
$H_{\text{S}}$ in determining transitions between different
subspaces of the dissipator. \ToAdd{The appearance of such
constraints in the time evolution is just the signature of the
(thermal) quantum Zeno effect, which gives rise to an inhibition
of the time evolution if the initial state of the system belongs
to a one-dimensional eigenspace of the dissipator, while gives
rise to a Zeno dynamics (evolution in the restricted subspace) if
the initial state belongs to a multiplet of the dissipator.}

{\bf\it Three-state system in a high-Temperature reservoir --- }
As a very special case, \ToAdd{we analyze the following example of
inhibition of the time evolution induced by temperature. Let us
consider a three-state system governed by the following
Hamiltonian:}
\begin{eqnarray}
H_{\text{S}} &= \sum_{l=1}^3 \nu_l \Projector{l} + \left( \Omega \KetBra{1}{2} + \mathrm{H.c.} \right)\,,\\
H_{\text{B}} &= \sum_k \omega_k a^\dag_k a_k\,,\\
H_{\text{I}} &= \left( \KetBra{2}{3} + \KetBra{3}{2} \right)
\otimes \sum_{k}g_k\left(a_k+a_k^\dag\right)\,.
\end{eqnarray}

\begin{figure}
\centering
\includegraphics[width=0.35\textwidth, angle=0]{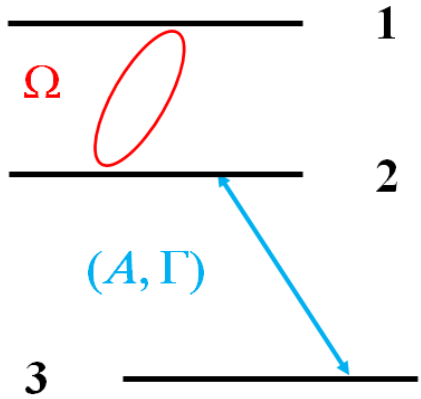} %
\caption{(Color online). Coupling scheme of the considered special
case: states $1$ and $2$ are coupled by some external field; $2$
and $3$ are coupled by the environment. The subspace $\Pi_1$ is
generated by $\Ket{1}$ and $\Ket{2}$,
while $\Pi_2$ coincides with $\Ket{3}$}. %
\label{fig:3_level_scheme}
\end{figure}

This model, which is represented in Fig.~\ref{fig:3_level_scheme},
has similarities with those analyzed in \cite{ref:Scala_LZ} and
\cite{ref:Militello_ThermalQZE}. Nevertheless, it differs from the
former because in the present case the Hamiltonian is time
independent, and it differs from the latter since here
counter-rotating terms are included in the system-bath
interaction.

On the basis of Eq.(\ref{eq:MasterEq_simplified_more}), at very
high temperature we obtain,
\begin{equation}%
\label{eq:MasterEq_simplified_more_special}%
\frac{\mathrm{d}\rho}{\mathrm{d}t} = {\cal G }\rho\,,
\end{equation}
with
\begin{equation}
{\cal G }\rho \approx -i\left[H_{\text{S}}, \rho \right] +
\alpha %
\left[ %
(  2 A\,\rho\,A - A^2\,\rho - \rho\,A^2 ) \right] %
\,,
\end{equation}
\begin{equation}
A = \KetBra{2}{3} + \KetBra{3}{2}\,.
\end{equation}

This master equation has been obtained under the previous
assumptions, i.e., high-temperature, band structure and intra-band
transitions forbidden.

The matrix representation of the superoperator ${\cal G}$ is the
following:
\newcommand{\iw}{\ii\omega}
\newcommand{\cp}{\alpha}
\begin{eqnarray}%
  \label{eq:MatrixDissipator}
  \nonumber
  {\cal G} = \\
  \nonumber
  \hskip-2.5cm \left(%
  \begin{array}{ccccccccc}
    0            & \ii\Omega^*   & 0             & -\ii\Omega    & 0           & 0              & 0             & 0              & 0          \\
    \ii\Omega    & -\cp-\iw_{12} & 0             & 0             & -\ii\Omega  & 0              & 0             & 0              & 0          \\
    0            & 0             & -\cp-\iw_{13} & 0             & 0           & 0              & 0             & 0              & 0          \\
    -\ii\Omega^* & 0             & 0             & -\cp-\iw_{21} & \ii\Omega^* & 0              & 0             & 0              & 0          \\
    0            & -\ii\Omega^*  & 0             & \ii\Omega     & -2\cp       & 0              & 0             & 0              & 2\cp       \\
    0            & 0             & 0             & 0             & 0           & -2\cp-\iw_{23} & 0             & 2\cp           & 0          \\
    0            & 0             & 0             & 0             & 0           & 0              & -\cp-\iw_{31} & 0              & 0          \\
    0            & 0             & 0             & 0             & 0           & 2\cp           & 0             & -2\cp-\iw_{32} & 0          \\
    0            & 0             & 0             & 0             & 2\cp        & 0              & 0             & 0              & -2\cp      \\
  \end{array}
  \right)\,,\\
\end{eqnarray}
where we have assumed the ordering $(\rho_{11}, \rho_{12},
\rho_{13}, \rho_{21}, \rho_{22}, \rho_{23}, \rho_{31}, \rho_{32},
\rho_{33})$, and exploited the notation $\omega_{ij} =
\nu_i-\nu_j$.

\ToAdd{The structure of this matrix is very similar to that we
have found in Ref~\cite{ref:Scala_LZ}. On the basis of the
analysis developed therein,} one easily deduces that for very
large $\alpha$ a Zeno Phenomenon occurs. In fact, \ToAdd{in the
$\alpha \rightarrow \infty$ limit (to better visualize this
situation, consider the complementary situation in which all the
other quantities tends toward zero),} the state $\KetBra{1}{1}$
turns out to be very close to an eigenvector of ${\cal G}$ and
therefore its survival probability approaches unity at every time:
the higher $\alpha$, the closer to unity is the survival
probability of $\KetBra{1}{1}$ at every time.

This is in line with the results in \cite{ref:Scala_LZ}, where no
secular approximation has been made, and generalizes the results
of \cite{ref:Militello_ThermalQZE}, where counter-rotating terms
are removed from the beginning in the interaction.

To reach the limit of very large $\alpha$, one needs a very high
temperature compared to $\omega_0$, meaning that the thermal
energy is supposed to be much larger than all the transition
frequencies, i.e., $\alpha \gg \omega_{13}, \omega_{23}$ (both
$\sim \omega_0$) and, {\it a fortiori}, $\alpha \gg \omega_{12}$.
Also $\alpha \gg \Omega$ must be satisfied.

\section{Conclusions}\label{sec:conclusions}

In this paper we have presented a \ToDelete{general} derivation of
Markovian master equations beyond the secular approximation which
is valid \ToAdd{for a class of physical systems satisfying}
\ToDelete{under} suitable hypotheses \ToDelete{concerning the
couplings induced by the bath}. In particular, it is important
that the energy levels of the system form two bands and that the
allowed transitions induced by the interaction with the bath are
only between states of different bands.

The structure of the master equation obtained exhibits some
important properties. First of all, it generates a completely
positive map, in spite of the fact that it is derived beyond the
secular approximation. Second, the structure of the dissipator is
independent from the Hamiltonian associated to the small system
and depends only on the operators involved in the system-bath
interaction. This is a consequence of the first property, since
summing up over all the jump operators associated to Bohr
frequencies of the system restores the complete form of the system
operator involved in system-bath coupling. Finally, on the basis
of the structure of the master equation it is easy to forecast the
incoming of Zeno phenomena at high temperature, which supports the
results obtained in some previous works.

\section{Acknowledgements}

MS acknowledges financial support from the EPSRC grant
EP/J014664/1.

\end{document}